\documentclass[oneside, twocolumn, prd, amsmath, amssymb, nofootinbib, superscriptaddress]{revtex4-2}

\usepackage{cases}
\usepackage{amsmath}
\usepackage{amssymb}
\usepackage{amsfonts}
\usepackage{amssymb}
\usepackage{dcolumn}

\usepackage{bm}
\usepackage{bbm}
\usepackage{graphicx}
\usepackage{xcolor}
\usepackage{array}
\usepackage{subfigure}
\usepackage{hyperref}
\usepackage{wasysym}
\usepackage{float}

\newcommand{\be}{\begin{equation}}
\newcommand{\ee}{\end{equation}}
\newcommand{\ba}{\begin{eqnarray}}
\newcommand{\ea}{\end{eqnarray}}

\newcommand{\gsim}{\mathrel{\hbox{\rlap{\lower.55ex \hbox {$\sim$}}
                   \kern-.3em \raise.4ex \hbox{$>$}}}}
\newcommand{\lsim}{\mathrel{\hbox{\rlap{\lower.55ex \hbox {$\sim$}}
                   \kern-.3em \raise.4ex \hbox{$<$}}}}

\hypersetup{colorlinks=true,
            breaklinks=true,
            pdfstartview=Fit,
            linkcolor=blue,
            citecolor=blue,
            urlcolor=blue}

\begin{document}
\title{Termination of Superradiance from a Binary Companion}

\author{Xi Tong}
\email{xtongac@connect.ust.hk}

\author{Yi Wang}
\email{phyw@ust.hk}

\author{Hui-Yu Zhu}
\email{hzhuav@connect.ust.hk}

\affiliation{Department of Physics, The Hong Kong University of Science and Technology, Clear Water Bay, Kowloon, Hong Kong, P.R.China}
\affiliation{The HKUST Jockey Club Institute for Advanced Study, The Hong Kong University of Science and Technology, Clear Water Bay, Kowloon, Hong Kong, P.R.China}

\begin{abstract}
We study the impact of a binary companion on black hole superradiance at orbital frequencies away from the gravitational-collider-physics (GCP) resonance bands. A superradiant state can couple to a strongly absorptive state via the tidal perturbation of the companion, thereby acquiring a suppressed superradiance rate. Below a critical binary separation, this superradiance rate becomes negative, and the boson cloud gets absorbed by the black hole. This critical binary separation leads to tight constraints on GCP. Especially, a companion with mass ratio $q>10^{-3}$ invalidates all GCP fine structure transitions, as well as almost all Bohr transitions except those from the $|\psi_{211}\rangle$ state. Meanwhile, the backreaction on the companion manifests itself as a torque acting on the binary, producing floating/sinking orbits that can be verified via pulsar timing. In addition, the possible termination of cloud growth may help to alleviate the current bounds on the ultralight boson mass from various null detections.
\end{abstract}

\maketitle

\section{Introduction}
Superradiance instability of ultralight bosons near a spinning black hole is a well-studied topic. The dissipative ergoregion of a Kerr black hole amplifies incoming waves and radiates out particles classically, a phenomenon widely known as superradiance \cite{zel1971generation,zel1972amplification,brito2020superradiance}. More interestingly, if the bosonic particle carries a non-zero mass, its mass barrier located at a Compton wavelength away from the ergoregion reflects the amplified wave back, thereby generating more particles \cite{Press:1972zz,Damour:1976kh,arvanitaki2010string}. These copiously produced bosons then condensate into clouds surrounding the black hole, with structures similar to those of an electron in the hydrogen atom \cite{arvanitaki2010string,arvanitaki2011exploring}.

Such a gravitational atom enjoys very rich phenomenology. For an isolated gravitational atom, the boson cloud quickly extracts the spin of black hole up to a saturation value. Observing black holes with spin higher than the saturation value then constrains superradiance and the boson properties \cite{Arvanitaki:2016qwi,Brito:2017zvb,Ng:2020ruv}. The boson cloud also emits monochromatic gravitational waves via pair annihilations \cite{arvanitaki2011exploring,yoshino2014gravitational}, which are potential targets for gravitational wave detectors such as LIGO and LISA. Null detection then puts bounds on the boson mass \cite{Palomba:2019vxe,LIGOScientific:2021jlr,LIGOScientific:2022lsr,Yuan:2022bem}. On the other hand, if the gravitational atom is in a binary system, more exciting physics kick in. For instance, a gravitational atom in a binary exhibits resonant transitions triggered by the orbital motion \cite{baumann2019probing,baumann2020gravitational,Baumann:2021fkf}, observable via gravitational wave probes and pulsar timing \cite{ding2021gravitational,tong2022gravitational}. These so-called Gravitational Collider Physics (GCP) \cite{baumann2020gravitational} transitions contain valuable information about the structure of the cloud as well as the properties of the boson. In addition, the mass quadrupole of the cloud induces orbital precession in an eccentric binary \cite{su2021probing}, observable from the gravitational wave or pulsar-timing signatures. At orbital separations close to the cloud radius, molecular structures emerge with distinctive observational signatures \cite{Ikeda:2020xvt,Ficarra:2021qeh,liu2021bh}.

However, we point out that, hidden in many of the phenomena above, is the crucial assumption of the existence of a boson cloud with total mass not far away from ($e.g.$, $1$-$3$ orders of magnitude smaller than) that of the black hole itself. In the case of a truly isolated black hole, superradiance guarantees the saturation of cloud growth, and the cloud can only deplete slowly via gravitational wave emission. In the presence of a binary companion, however, the stability of cloud is far from obvious. For example, it is known that the resonant GCP transitions may deplete the cloud efficiently \cite{baumann2019probing,Berti:2019wnn,Takahashi:2021eso}, while ionization effects may evaporate the cloud \cite{Wong:2020qom,Baumann:2021fkf,Takahashi:2021yhy}.

\begin{figure}[h!]
	\centering
	\includegraphics[width=8cm]{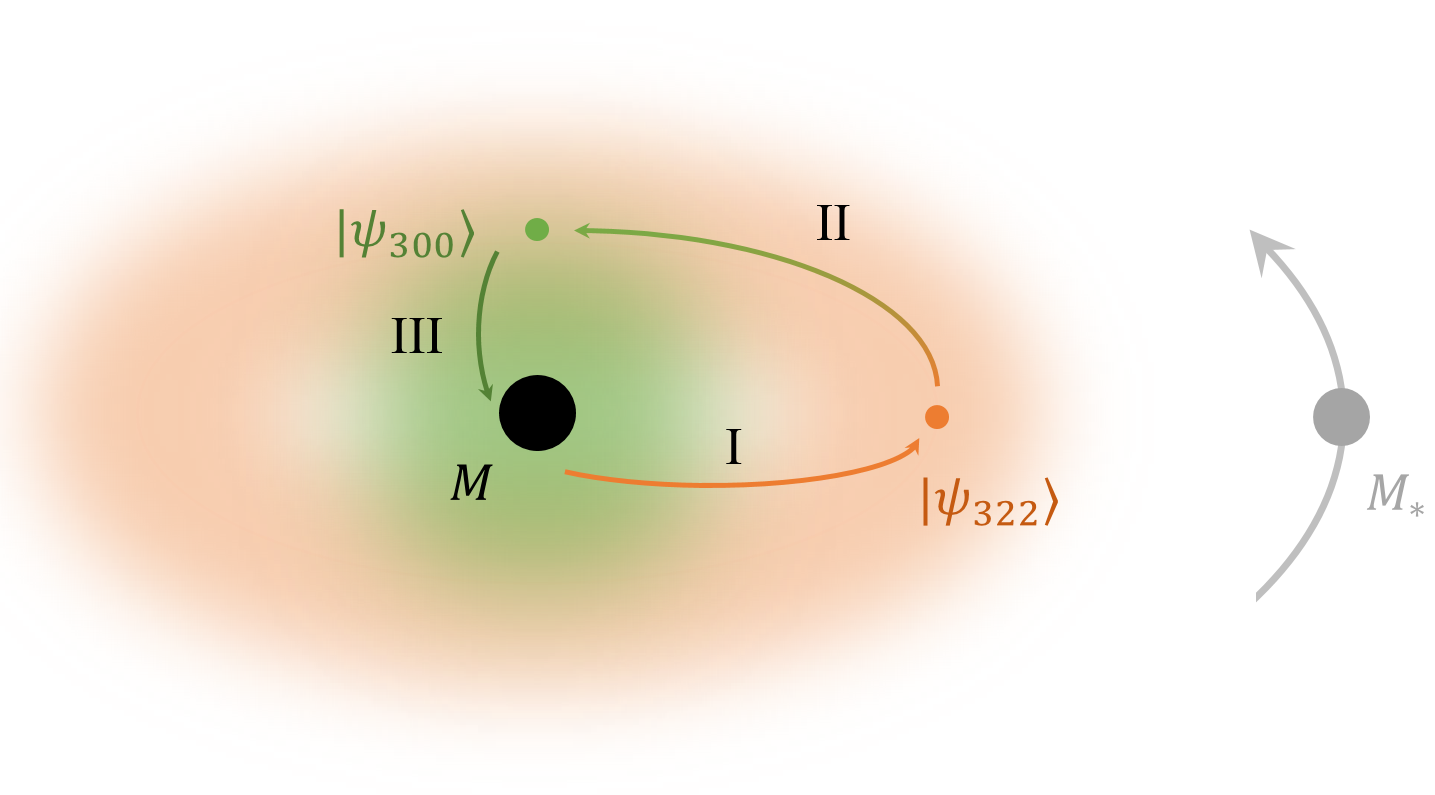}
	\caption{The life cycle of a boson near a rotating black hole in binary. I: The boson is produced via black hole superradiance instability. II: The tidal perturbation of the binary companion (gray) turns the boson from a superradiant state ($|\psi_{322}\rangle$) to an absorptive state ($|\psi_{300}\rangle$) in an \textit{off-resonance} fashion. III: The boson is quickly reabsorbed by the black hole. This whole process contributes negatively to the growth rate of the boson cloud, thereby suppressing superradiance.}\label{figFancy}
\end{figure}

In this work, we highlight the fact that the superradiant modes ($e.g.,~|\psi_{322}\rangle$) can overlap with the dangerous absorptive modes ($e.g.,~|\psi_{300}\rangle$) via the tidal perturbation of the binary companion. Depending on the binary separation, the superradiance growth rate may be suppressed, and even become negative if the companion is too close (see FIG.~\ref{figFancy} for a cartoon illustration). Consequently, the cloud cannot exist when the binary separation is below a critical distance. This dramatically affects the gravitational atom phenomenology. First, the parameter space of certain GCP transitions becomes constrained. Second, we may still observe orbital period derivative changes due to the backreaction of cloud absorption, which produces floating/sinking orbits \textit{away} from the GCP resonance bands. Third, the bound on boson masses from various null detections may be alleviated, due to unknown companion objects depleting the cloud.

This paper is organized as follows. In Sect.~\ref{BHSTPReviewSection}, we first review the basic ingredients of black hole superradiance and tidal perturbation theory. Then in Sect.~\ref{SvASection}, we compute the suppressed superradiance rate due to the gravitational perturbation of the binary, and determine the critical distances below which superradiance is turned off. We discuss the implications for the gravitational atom phenomenology in Sect.~\ref{ObsSection}. At last, we conclude in Sect.~\ref{ConclusionSection}. We set $G=\hbar=c=1$ throughout the paper and our conventions and notations largely follow from \cite{baumann2020gravitational,ding2021gravitational}.

\section{The gravitational atom and tidal perturbations}\label{BHSTPReviewSection}
In this section, we begin with a lightning review of black hole superradiance instability and tidal perturbation theory. Consider an ultralight (scalar) boson field with mass $\mu$ around a Kerr black hole with mass $M$ and dimensionless spin parameter $\tilde{a}$. The bosons generated by the Kerr black hole condensate into clouds and form a bound state together with the black hole. This spectrum of this gravitational atom is governed by the Klein-Gordon equation in Kerr spacetime:
\begin{equation}
	\left(g^{\alpha\beta}\,\nabla_\alpha\,\nabla_\beta-\mu^2\right)\Phi=0~. 
	\label{KG}
\end{equation}
If the gravitational fine structure constant $\alpha\equiv G M \mu\ll 1$ is perturbatively small, the cloud is mostly non-relativistic and we can factor out a rest-mass dynamical phase from the boson field,
\begin{equation}
	\Phi\equiv \frac{1}{\sqrt{2\mu}}e^{-i\mu t}\psi+\text{c.c.}~.\label{NRApprox}
\end{equation}
Using the Boyer-Lindquist coordinates, we insert (\ref{NRApprox}) into the Klein-Gordon equation and expand in powers of $\alpha$, the resulting equation takes a Schr\"{o}dinger form at leading order,
\begin{equation}
	i\partial_t\psi(t,\mathbf{r})=H_0\psi(t,\mathbf{r})~,~H_0\equiv -\frac{1}{2\mu}\partial_{\mathbf{r}}^2+V(r)~,
\end{equation}
where the potential $V(r)=-\frac{\alpha}{r}+\mathcal{O}(\alpha^2)$ resembles that of a hydrogen atom. Solving the system with in-going boundary condition at the black hole outer horizon yields the quasi-stationary bound states, which at large distance ($r\gg M$) approximate to
\begin{equation}
	\psi_{nlm}(\bm{r})=R_{nl}(r)\,Y_{lm}(\theta,\phi)\,e^{-i\,(\omega_{nlm}-\mu)t}~, 
	\label{ansatz} 
\end{equation}
where $R_{nl}$ is the hydrogen radial function and $n,l,m$ are the usual principle, angular and magnetic quantum numbers. The key difference from that of hydrogen atom is that the eigenvalues of boson clouds are in general complex (hence the instability),
\begin{equation}
	\omega_{nlm}=E_{nlm}+i\,\Gamma_{nlm}~,
\end{equation}
due to an absorptive horizon and dissipative ergoregion. The detailed form of the energy levels and decay widths can be found from inspecting the higher-order corrections of the Kerr geometry. Up to hyperfine splittings, the energy levels are given by \cite{baumann2019probing,baumann2019spectra}
\begin{equation}
	\begin{aligned}  
	E_{nlm}&=\mu\left(1-\frac{\alpha^2}{2\,n^2}-\frac{\alpha^4}{8\,n^4}-\frac{(3\,n-2\,l-1)\,\alpha^4}{n^4\,(l+1/2)}\right)\\
	&+\frac{2\,\tilde{a}\,m\,\alpha^5}{n^3\,l\,(l+1/2)\,(l+1)}+\mathcal{O}(\alpha^6)~.
	\end{aligned}
\end{equation}
The decay widths are given at $\alpha\lesssim 0.4$ via the Detweiler approximation\footnote{Note that numerics shows a factor-of-two uncertainty in the Detweiler approximation, although this is not important for our purpose. See a recent refinement in \cite{bao2022improved}} \cite{detweiler1980klein},
\begin{equation}
	\begin{aligned}
		\Gamma_{n00}&= -\frac{4}{n^3}(1+\sqrt{1-\tilde{a}^2})\,\mu\,\alpha^5\\ 
		\Gamma_{nlm}&=2\tilde{r}_+\,C_{nl}\,g_{lm}(\tilde{a},\alpha,\omega)\,(m\,\Omega_H-\omega_{nlm})\,\alpha^{4l+5}. 
	\end{aligned}
\end{equation}
Here $\tilde{r}\equiv r/M$, $\tilde{r}_+\equiv 1+\sqrt{1-\tilde{a}^2}$ and $\Omega_H\equiv\tilde{a}/[2\,M^2\,(1+\sqrt{1+\tilde{a}^2})]$ is the angular velocity of the outer horizon. The definition of $C_{nl}$ and $g_{lm}$ are given as. 
\begin{equation}
	C_{nl}\equiv\frac{2^{4l+1}(n+l)!}{n^{2l+4}(n-l-1)!}\,\left[\frac{l!}{(2l)!\,(2l+1!)}\right]^2, 
\end{equation} 
\begin{equation}
	g_{lm}(\tilde{a},\alpha,\omega)\equiv\prod_{k=1}^{l}\left(k^2\,(1-\tilde{a}^2)+(\tilde{a}m-2\tilde{r}_+ M\omega)^2\right). 
\end{equation}

For an isolated black hole, the occupation number of superradiant states ($\Gamma_{nlm}>0$) grows at a timescale $T_{nlm}^{(\text{grow})}=\Gamma_{nlm}^{-1}\propto \alpha^{-(4l+5)}$ until saturation. In contrast, states with $\Gamma_{nlm}<0$ will be absorbed into the black hole at a timescale $T_{nlm}^{(\text{absorb})}=\Gamma_{nlm}^{-1}\propto \alpha^{-(4l+5)}$. After saturation, the superradiant states constantly emit monochromatic gravitational waves and deplete at a much longer time scale $T^{(\text{deplete})}\propto \alpha^{-(4l+10)}$.

Now consider inducing a binary companion of mass $M_*$. At leading order, we can understand the influence of a binary companion as generating a time-periodic tidal field around the gravitational atom. In the Fermi normal coordinates, the Newtonian potential of the binary companion can be expanded using spherical harmonics as
\begin{align}
		\nonumber V_*=&-\alpha \,q\,\sum_{l_*\geqslant 2}\sum_{|m_*|\leqslant l_*}\,\mathcal{E}_{l_*m_*}(\iota_*,\varphi_*)\,Y_{l_*m_*}(\theta,\phi)\\
		&\times\left(\frac{r^{l_*}}{R_*^{l_*+1}}\,\Theta(R_*-r)+\frac{R_*^{l_*}}{r_*^{l_*+1}}\,\Theta(r-R_*)\right)~.
\end{align}
Here $q=M_*/M$ is the mass ratio of the companion and the gravitational atom, $R_*$ is the binary separation, and $\iota_*, \varphi_*$ are the inclination angle and true anomaly, respectively. $\Theta(x)$ is the Heaviside step function. The tidal moment $\mathcal{E}$ can be computed from the geometry of the binary. For equatorial orbits, the quadrupole contribution simplifies considerably to
\begin{align}
		\mathcal{E}_{2\mp2}=\frac12\sqrt{\frac{6\pi}{5}}e^{\pm2i\varphi_*},\mathcal{E}_{2\mp1}=0,\mathcal{E}_{20}=-\sqrt{\frac\pi5}~.
		\label{tidal}
\end{align}
The tidal field of the binary companion decays as $R_*^{l_*+1}$ for the $l_*$-th multipole moment. Although $V_*$ is small for large separations, it introduces weak mixings between energy eigenstates of $H_0$. This overlap of two $H_0$ eigenstates is calculated as
\begin{align}
		\nonumber \langle \psi_{n'l'm'}|V_*(t)|\psi_{nlm}\rangle\equiv&(-1)^{m'+1}\alpha\, q\\
		&\sum_{l_*,m_*}\mathcal{E}_{l_*m_*}(t)\mathcal{G}^{l'l_*l}_{-m'm_*m}\,I_r, 
\label{overlap}
\end{align}
Here $\mathcal{G}^{l'l_*l}_{-m'm_*m}$ is the Gaunt integral,
\begin{equation}
	\mathcal{G}_{-m'm_*m}^{l'l_*l}=\int{\rm{d}}\Omega Y_{l'-m'}(\theta,\phi)Y_{l_*m_*}(\theta,\phi)Y_{lm}(\theta,\phi)~,
	\label{Gaunt}
\end{equation}
which implicitly imposes a set of selection rules,
\begin{equation}
	\begin{aligned}
		&-m'+m_*+m=0\\
		&l+l_*+l'=2p,\text{ for } p\in\mathbb{Z}\\
		&|l-l'|\leqslant l_*\leqslant l+l'~.
	\end{aligned}\label{SelectionRules}
\end{equation}
And $I_r$ is the radial integral,
\begin{equation}
	\begin{aligned}
	I_r&=\int_0^{R^*} r^2{\rm{d}}rR^*_{n'l'}(r)R_{nl}(r)\frac{r^{l_*}}{R_*^{l_*+1}}\\
	&+\int_{R^*}^\infty r^2{\rm{d}}rR^*_{n'l'}(r)R_{nl}(r)\frac{R_*^{l_*}}{r_*^{l_*+1}} ~.
\end{aligned}
\end{equation}
The radial integral is dominated by the first term if the companion is far outside the cloud, $i.e.$, $R_*\gg n^2 r_1$, where $r_1\equiv(\mu \alpha)^{-1}=M/\alpha^2$ is the Bohr radius.

The dynamics of the system is then determined from solving the full Hamiltonian $H\equiv H_0+V_*$ with orbital backreaction. It is widely known that the periodic tidal perturbation $V_*$ can trigger atomic resonances when the orbital frequency matches the energy difference of two $H_0$-eigenstates. These GCP resonances exhibit distinctive floating/sinking features in their orbital evolutions, which can be observed via multiple messengers \cite{tong2022gravitational}.


\section{Termination of superradiance}\label{SvASection}
One might think that when the binary separation is much larger than the cloud radius, $i.e.$, $R_*\gg r_1\sim (\mu\alpha)^{-1}$, the gravitational perturbation of the binary companion ($\langle V_*\rangle \sim \alpha r_1^2/R_*^{3}\ll \mu \alpha^2$) may be too small to affect the cloud dynamics, which is typically of order $\langle H_0\rangle \sim \mu \alpha^2$. However, one must recall that the superradiance rate, which governs cloud formation, is also small. For instance, the $|\psi_{322}\rangle$ state has a superradiance rate $\Gamma_{322}\sim \mu\alpha^{12}$. Therefore, it might be the case that the cloud formation process can be dramatically influenced by the presence of a binary companion. Indeed, we will see that the tidal perturbation of the binary companion typically suppresses the superradiance rate, and may even terminate superradiance and prevent the cloud from ever growing up.

\subsection{The adiabatic case}
We start with the perturbed Hamiltonian $H=H_0+V_*$ of the boson. Its matrix element is given by
\begin{equation}
	\langle \psi_{n'l'm'}|H|\psi_{nlm}\rangle=\omega_{nlm}\delta_{n'n}\delta_{l'l}\delta_{m'm}+\langle \psi_{n'l'm'}|V_*|\psi_{nlm}\rangle~. 
	\label{H} 
\end{equation}
The diagonal terms are dominated by the eigenvalues of the free Hamiltonian while the off-diagonal terms are led by the gravitational level mixings due to the companion. The level mixings couple two states together whenever the selection rule allows. In particular, the superradiant states can be coupled to the absorptive states, bringing negative contributions to the imaginary part of the frequencies. Most dangerous of all absorptive states are the spherically symmetric $|\psi_{n00}\rangle$ states, since they decay the fastest. Hence we expect the leading order correction to the superradiant rate of $|\psi_{nlm}\rangle$ to come from its mixing with $|\psi_{n00}\rangle$ (if the selection rule allows). Therefore, we will focus on a two-state subspace $\{|1\rangle,|2\rangle\}$, where $|1\rangle$ denotes a superradiant state $|\psi_{nlm}\rangle$ and $|2\rangle$ denotes a highly absorptive state such as $|\psi_{n00}\rangle$. In matrix form, the perturbed Hamiltonian reads
\begin{align}
	H=   \left(
	\begin{matrix}
		\omega_1+V_{11} & V_{12} \\
		V_{21} & \omega_2+V_{22}
	\end{matrix}
	\right)
	\equiv
	\left(
	\begin{matrix}
		\bar{E}_1+i\Gamma_1 & \eta^* \\
		\eta & \bar{E}_2+i\Gamma_2
	\end{matrix}
	\right)~,
	\label{perturbedHamiltonian}
\end{align} 
where we have denoted $\eta\equiv V_{21}$, $\bar{\omega}_i\equiv \omega_i+V_{ii}$ and $\bar{E}_i\equiv E_i+V_{ii}$, $i=1,2$. Notice that both $\bar{E}_i(t)$ and $\eta(t)$ implicitly depend on time through the binary orbital motion. Thus in the adiabatic limit $|\frac{\dot\eta}{\bar E_i^2}|,|\frac{\dot{\bar{E}}_i}{\bar E_i^2}|\ll 1$, the state of the cloud $c_i(t)\equiv\langle \psi_i|\psi(t)\rangle$ can be solved using the WKB approximation,
\begin{equation}
	c_i(t)=C_{i+}e^{-i\int\lambda_+ dt}+C_{i-}e^{-i\int \lambda_- dt}~,~i=1,2~,\label{naiveWKB}
\end{equation}
where $\lambda_\pm$ are the two instantaneous eigenvalues of the perturbed Hamiltonian (\ref{perturbedHamiltonian}),
\begin{align}
	\nonumber \lambda_\pm\equiv
	&\frac{\bar\omega_1+\bar\omega_2}{2}\pm\sqrt{|\eta|^2+\left(\frac{\bar\omega_1-\bar\omega_2}{2}\right)^2}\\
	\simeq&\left\{
		\begin{array}{lr}
			 \bar E_1+\frac{|\eta|^2}{\bar E_1-\bar E_2}+i\left[\Gamma_1-\frac{\Gamma_1-\Gamma_2}{(\bar E_1-\bar E_2)^2}|\eta|^2\right] \text{, } +&\\\\
			 \bar E_2+\frac{|\eta|^2}{\bar E_2-\bar E_1}+i\left[\Gamma_2-\frac{\Gamma_2-\Gamma_1}{(\bar E_1-\bar E_2)^2}|\eta|^2\right] \text{, }- &
		\end{array}
	\right.
\end{align}
The approximation is made under the limit $|\bar E_{1,2}|\gg|\eta|,|\Gamma_{1,2}|$, which is always valid given the superradiance context. Thus we see that $\lambda_+$ ($\lambda_-$) represents the corrected frequency of the quasi-stationary state $|1\rangle$ ($|2\rangle$). In particular, their imaginary parts are modified. Since we have assumed $\Gamma_1\geqslant0$ and $\Gamma_2<0$, the effective superradiance rate is now suppressed:
\begin{equation}
	\tilde{\Gamma}_1\equiv \Gamma_1+\Delta \Gamma_1,~\Delta\Gamma_1\simeq-\frac{\Gamma_1-\Gamma_2}{(\bar E_1-\bar E_2)^2}|\eta(R_*)|^2<0~.
	\label{correction}
\end{equation}

This suppression term in the effective superradiance rate mainly depends on the distance $R_*$ between the binary. If the absorptive state $|2\rangle$ happens to be a highly dangerous state such as $|\psi_{n00}\rangle$, the corresponding suppression can be significant if $R_*$ is not large. For instance, consider the case $n=3,l=m=2$. The fine-structure splitting is
\begin{equation}
	E_{322}-E_{300}\simeq 0.04~ \text{s}^{-1}\left(\frac{\alpha}{0.1}\right)^{5}\left(\frac{M}{10M_{\odot}}\right)^{-1}~.
\end{equation}
The superradiance/absorption rates at maximal black hole spin ($\tilde{a}=1$) are
\begin{align}
	\Gamma_{322}&\simeq 3\times 10^{-13}~ \text{s}^{-1}\left(\frac{\alpha}{0.1}\right)^{13}\left(\frac{M}{10M_{\odot}}\right)^{-1}~,\label{322Gamma}\\
	\Gamma_{300}&\simeq -3\times 10^{-3}~ \text{s}^{-1}\left(\frac{\alpha}{0.1}\right)^{5}\left(\frac{M}{10M_{\odot}}\right)^{-1}~.
\end{align}
Then at a binary separation $R_*=10^5 M$, the size of the level mixing is
\begin{equation}
	|\eta(R_*)|\simeq 2\times 10^{-7}~ \text{s}^{-1}\left(\frac{\alpha}{0.1}\right)^{-3}\frac{q}{0.2}\left(\frac{M}{10M_{\odot}}\right)^{-1}~.
\end{equation}
This results in a correction to the superradiance rate
\begin{align}
	\nonumber \Delta \Gamma_{322}\simeq& -7\times 10^3 \frac{ q^2}{\alpha^{10}}\frac{M^5}{R_*^6}\\
	\simeq& -0.6\times 10^{-13}~ \text{s}^{-1}\left(\frac{\alpha}{0.1}\right)^{-10}\left(\frac{q}{0.2}\right)^2\left(\frac{M}{10M_{\odot}}\right)^{-1}~,\label{322DeltaGamma}
\end{align}
which already gives $\tilde{\Gamma}_{322}\sim\frac{4}{5}\Gamma_{322}$, $i.e.$, a reduction of $20\%$. As one expects from the quadrupole tidal perturbation, $\Delta \Gamma_{322}$ grows as $R_*^{-6}$ as the binary separation decreases. Therefore, inside a critical distance $R_*<R_{*,c}$ defined from the condition
\begin{equation}
	\tilde{\Gamma}_{nlm}(R_{*,c})=\Gamma_{nlm}+\Delta \Gamma_{nlm}(R_{*,c})\equiv0~,\label{criticalDistanceDef}
\end{equation}
the effective superradiance rate becomes negative, and the mode $|\psi_{nlm}\rangle$ can no longer grow even at maximal black hole spin. For the $|\psi_{322}\rangle$ example, the critical distance is easily read out from (\ref{322Gamma}) and (\ref{322DeltaGamma}):
\begin{equation}
	R_{*,c}(322)\simeq 10^6~\text{km} \left(\frac{\alpha}{0.1}\right)^{-23/6}\left(\frac{q}{0.2}\right)^{1/3}\frac{M}{10M_{\odot}}~.
\end{equation}

Before moving on to further discussions, we would like to make a few comments on suppressed superradiance and the critical distance.
\begin{enumerate}
	\item[$\bullet$] First, it is interesting to see the interplay of UV and IR effects here. The imaginary part of the eigenvalues of the free Hamiltonian $H_0$ can be understood as an UV effect, which is relevant near the black hole horizon scale $r_{UV}\sim M$. Yet its correction involves IR effects which has nothing to do with the black hole geometry. These co-rotating bosons produced via superradiant scattering leak to an IR scale $r_{IR}\sim (\mu\alpha)^{-1}$, and are turned into non-/counter-rotating bosons by the tidal perturbation of the binary companion, which then get absorbed by the black hole again in the UV (see again FIG.~\ref{figFancy}).
	\item[$\bullet$] Second, if the binary separation is shorter than the critical distance $R_{*,c}(nlm)$, there can be two scenarios. If the cloud has not been formed, yet the black hole spin appears to be capable of superradiance ($\Gamma_{nlm}>0$), superradiance will be shut off. No $|\psi_{nlm}\rangle$-cloud will be produced, and black hole spin cannot drop below the saturation value for modes with magnetic quantum number $m$. This is because, unlike Hawking radiation, superradiance is not spontaneous and can only amplify a given cloud state. But if such a cloud state is easily depleted, there is no seed particles left for amplification, hence no extraction of black hole spin. On the other hand, if the cloud has already been formed, it will decay slowly at the rate $|\tilde{\Gamma}_{nlm}|$. Like GCP resonances, this process generates backreaction to the binary orbit and can be observed. More discussions on this possibility will be elaborated in Sect.~\ref{OffResObservability}.
	\item[$\bullet$] Third, the superradiance termination we are dealing with here is similar but not identical to resonant depletion scenarios considered in the literature \cite{baumann2019probing,Berti:2019wnn,Takahashi:2021eso}. Specifically, we need no requirement on the orbital frequency or orbit direction (co-rotating/counter-rotating). In fact, we will mostly stay away from the GCP resonance bands and analyze the viability of GCP resonances considering superradiance termination in Sect.~\ref{GCPConstraintSection}.
\end{enumerate}

\subsection{Beyond adiabaticity and multiple states}\label{FullCDSection}
The above calculations are performed under the assumption of adiabaticity. However, when the orbital frequency is high, the naive WKB approximation (\ref{naiveWKB}) breaks down and we need a better treatment of the perturbed Hamiltonian $H$. For simplicity, we will specialize to the case of equatorial circular orbits. The tidal moments thus reduce to
\begin{equation}
	\mathcal{E}_{l_*m_*}=\frac{4\pi}{2l_*+1}Y_{l_*m_*}(\pi/2,\varphi_*(t))=e_{l_*m_*} e^{i m_* \varphi_*(t)}~,
\end{equation}
where
\begin{equation}
	e_{l_*m_*}\equiv \sqrt{\frac{4\pi}{2l_*+1}\frac{(l_*-m_*)!}{(l_*+m_*)!}}P_{l_*}^{m_*}(0)
\end{equation}
and $P_{l_*}^{m_*}(x)$ is the associated Legendre polynomial. This $e^{i m_* \varphi_*(t)}$ time dependence is inherited by the off-diagonal term $\eta(t)$. Therefore, we follow the standard procedure and perform a time-dependent unitary transformation to go into the co-rotating frame \cite{takagi1991quantum,baumann2020gravitational,Ikeda:2020xvt},
\begin{align}
	\nonumber&H_D=U(t)^\dagger (H(t)-i\partial_t) U(t),\\
	&\text{with  }U(t)\equiv e^{-i\varphi_*(t) L_z}~,
\end{align}
where $L_z$ is the cloud angular momentum operator along the spin axis. In component form, the Hamiltonian in the co-rotating frame reads
\begin{equation}
	H_D=\left(
	\begin{matrix}
		\bar{E}_1+i\Gamma_1-m_1 \dot\varphi_* & |\eta| \\
		|\eta| & \bar{E}_2+i\Gamma_2-m_2 \dot\varphi_*
	\end{matrix}
	\right)~.
\end{equation}
Notice that we have applied the selection rule (\ref{SelectionRules}) and set $m_*=m_2-m_1$. Now the fast oscillations in $H$ are eliminated and we are left with a co-rotating frame Hamiltonian $H_D$ that varies slowly in time only through $R_*(t)$. Going through the same procedure as before, we obtain the correction to the superradiance rate of $|1\rangle$:
\begin{equation}
	\Delta \Gamma_1\simeq -\frac{\Gamma_1-\Gamma_2}{\left[\bar E_1-\bar E_2-(m_1-m_2)\dot\varphi_*(R_*)\right]^2}|\eta(R_*)|^2~.\label{2stateDeltaGamma}
\end{equation}
Clearly, in the adiabatic limit, $|\frac{\dot{\varphi}}{E_i}|\ll 1$, the above expression reduces to (\ref{correction}). The impact on superradiance becomes severe when the denominator vanishes. This corresponds to the resonant depletion scenario \cite{baumann2019probing,Berti:2019wnn,Takahashi:2021eso}, where the cloud state $|1\rangle$ largely mixes into $|2\rangle$ and is quickly absorbed into the black hole. Since we focus more on the off-resonance scenario, the denominator will typically be dominated either by $\bar{E}_1-\bar{E}_2$ (the adiabatic case) or by $(m_1-m_2)\dot\varphi$ (the diabatic case).

Since both the tidal perturbations and the instability rates are small compared to the energy levels, the two-state result (\ref{2stateDeltaGamma}) can be readily generalized to that of multiple states. To leading order in $|\eta_{ij}|$, the impact on superradiance is just a simple summation,
\begin{equation}
	\Delta \Gamma_1\simeq -\sum_{i=n'l'm'}\frac{\Gamma_1-\Gamma_i}{\left[\bar E_1-\bar E_i-(m_1-m_i)\dot\varphi_*(R_*)\right]^2}|\eta_{1i}(R_*)|^2~,\label{mulitstateDeltaGamma}
\end{equation}
where $\eta_{1i}$ contains a sum of tidal moments $\mathcal{E}_{l_*,m_i-m_1}$. The critical distance is again determined by requiring $\tilde\Gamma_1=\Gamma_1+\Delta \Gamma_1=0$. Due to the $R_*$ dependence in the denominator, there may be multiple solutions of $R_{*,c}$ when entering/exiting a resonance. We are more interested in the off-resonance solutions. 

\begin{figure}[h!]
	\centering
	\includegraphics[width=8.5cm]{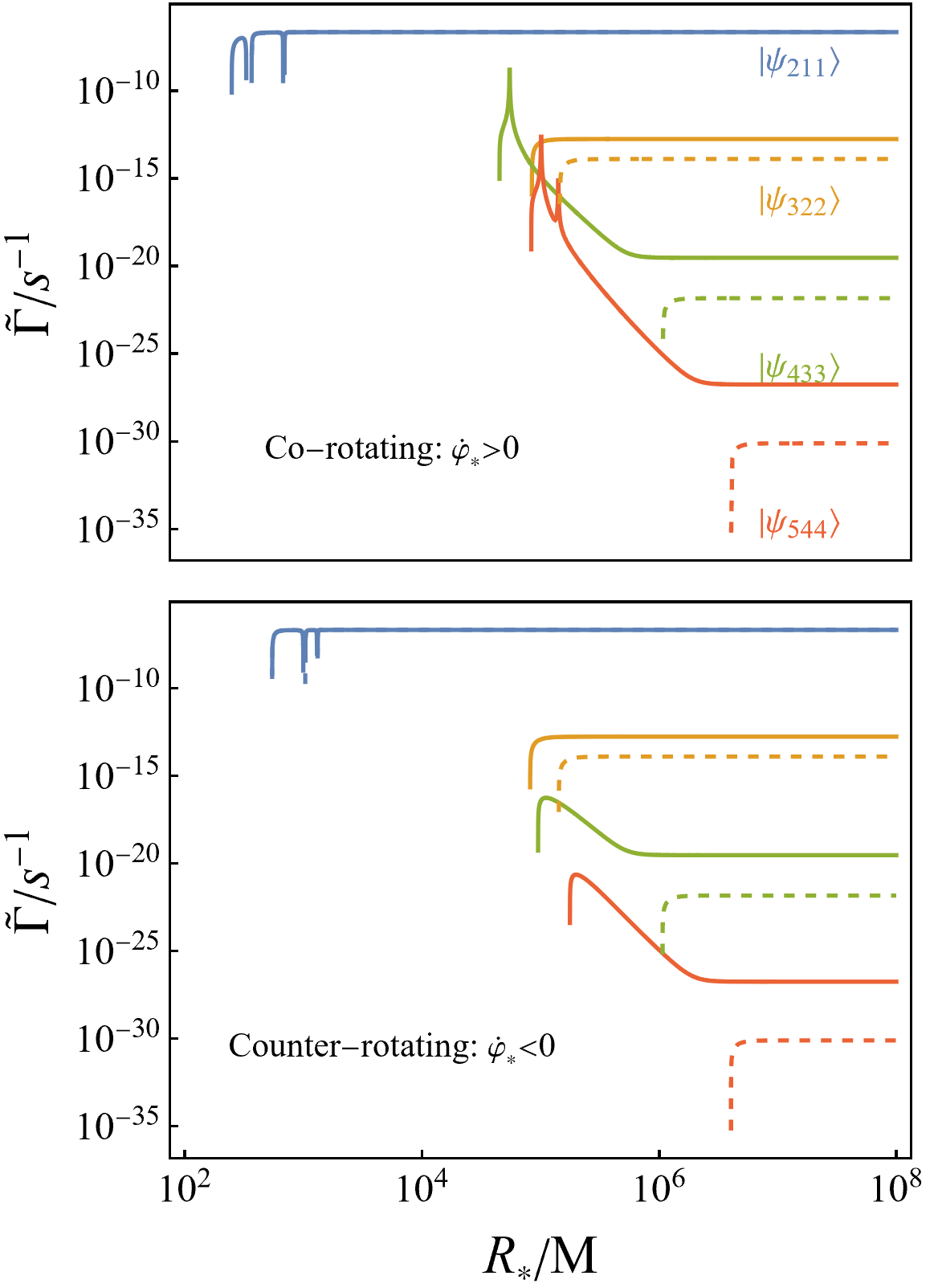}
	\caption{The effective superradiance rate $\tilde\Gamma_{nlm}$ as a function of the binary separation $R_*$ for co-rotating orbits (upper panel) and counter-rotating orbits (lower panel). The solid lines correspond to the maximal black hole spin $\tilde{a}=1$, while the dashed lines correspond to the previous saturation black hole spin $\frac{4(m-1)\alpha}{(m-1)^2+4\alpha^2}$ for the $m$-th state ($e.g.$, $\tilde{a}=\frac{2\alpha}{1+\alpha^2}$ when considering the $m=3$ state). The other parameters are chosen to be $\alpha=0.1$, $M=10M_{\odot}$ and $q=0.2$. It is clear that the effective superradiance rate reduces to the Detweiler value at large binary separations, while dropping below zero generically at a finite critical distance.}\label{figGammaEff}
\end{figure}

In FIG.~\ref{figGammaEff}, we have plotted the binary-separation-dependence of the effective superradiance rate $\tilde\Gamma_{nlm}(R_*)$ for different states and black hole spins. It can be seen that the effective superradiance rate $\tilde\Gamma_{nlm}$ tends to the Detweiler value $\Gamma_{nlm}$ at large binary separations, since the tidal perturbations become negligible in this limit. As we decrease the binary separation, $\tilde\Gamma_{nlm}$ quickly drops to zero near a critical distance, where superradiance is terminated due to mixing into absorptive states. The sharp peaks and valleys are caused by GCP resonances, where the mixing effect is non-perturbatively large. Interestingly, when the black hole spin is high ($e.g.$, $\tilde{a}=1$), the high-$l$ states such as $|\psi_{433}\rangle$ and $|\psi_{544}\rangle$ receive \textit{positive} enhancement to the superradiance rate first, before \textit{negative} suppression terms take over and terminate superradiance as $R_*$ decreases. This is because of their mixings into low-$l$ states such as $|\psi_{411}\rangle$ and $|\psi_{522}\rangle$, which possess larger positive growth rates. Yet as the black hole spin drops below the threshold for superradiating these low-$l$ states, the superradiance rate $\tilde\Gamma_{nlm}$ for high-$l$ states can no longer benefit from the mixings, and thus monotonically drop to zero as $R_*$ decreases, as indicated by the dashed lines in FIG.~\ref{figGammaEff}.

The critical distance can be solved numerically from (\ref{criticalDistanceDef}) in the case of multiple states beyond adiabaticity. As mentioned before, we focus on systems with an orbital frequency away from the GCP resonance bands, because such a configuration occupies the most proportion of the binary lifetime and are more typical statistically. Henceforth, it will be convenient to perform a ``Wick'' rotation and replace the denominator in (\ref{mulitstateDeltaGamma}) by
\begin{align}
	\nonumber&\frac{1}{\left[\bar E_1-\bar E_i-(m_1-m_i)\dot\varphi_*(R_*)\right]^{2}}\\
	&\to\frac{1}{(\bar E_1-\bar E_i)^2+[(m_1-m_i)\dot\varphi_*(R_*)]^2}~.
\end{align}
This modification removes the resonance poles, but it keeps the off-resonance physics mostly unaltered. Since the tidal-perturbation-theory calculations are under the assumption of $R_*>r_n$, the resulting critical distance must be greater than the cloud radius,
\begin{equation}
	R_{*,c}(nlm)>r_n=n^2 r_1\text{ for consistency}~.\label{consistencyConstraint}
\end{equation}

\begin{figure}[h!]
	\centering
	\includegraphics[width=9cm]{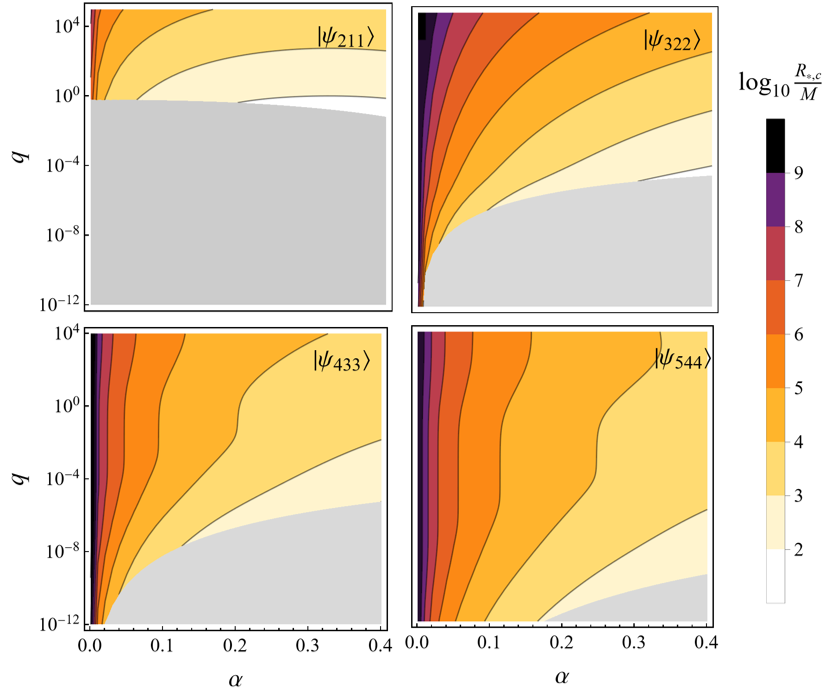}
	\caption{The critical distance $R_{*,c}$ (in units of the horizon size) as a function of $\alpha$ and $q$ for different cloud states. The gray region is excluded by the consistency constraint (\ref{consistencyConstraint}). The black hole spin is chosen to be maximal, $\tilde{a}=1$. Therefore, below the critical distance, no net superradiance is present and no existing boson cloud can last for a period longer than its original typical growth time.}\label{figRc}
\end{figure}

Under this constraint, we solve (\ref{criticalDistanceDef}) for the critical distance and plot its dependence on $\alpha$ and $q$ in FIG.~\ref{figRc}. Because we have chosen a maximal black hole spin, the resulting critical distance is the minimal requirement for cloud stability. This means superradiance of a given state $|\psi_{nlm}\rangle$ will be terminated completely below $R_*<R_{*,c}(nlm)$, and an existing cloud will decay quickly at a rate comparable to its original growth rate. From FIG.~\ref{figRc}, we see that for fixed $\alpha$ and $M$ (hence fixed boson mass), the critical distance decreases with a smaller mass ratio. This is in agreement with intuition, since a lighter (heavier) binary companion is expected to have weaker (stronger) gravitational perturbations on the cloud dynamics, resulting in a broader (narrower) safe region. Cloud states with higher $l$ appear to be less stable in the presence of a binary companion, as their critical distances are larger. At last, we stress that as an off-resonance quantity, the critical distance solved in this way does not depend on the orbit orientation, which matters only for the triggering of a GCP resonance.

\section{Observational consequences}\label{ObsSection}
Given the potential threat of a binary companion to the boson cloud, how does this superradiance termination effect influence the phenomenology of the gravitational atom? In this section, we will briefly examine three of its main consequences.

\subsection{Implications for GCP transitions}\label{GCPConstraintSection}
First, the termination of superradiance poses a constraint on observable GCP transitions. This is natural since GCP is based on resonant cloud transitions. If there is no cloud when the binary enters the resonance band, there is certainly no cloud transition, and no observable signal. Therefore, roughly speaking, in order to have successful GCP transitions, we must require that the binary separation at resonance to be greater than the critical distance of the initial cloud state\footnote{Notice that both sides of (\ref{GCPAllowedCondition}) depend on $\alpha$ and $q$. Thus this is effectively a constraint on the relative sizes of three mass parameters (namely, $\mu, M, M_*$) in the system.},
\begin{equation}
	R_{*,r}(nlm\to n'l'm')>R_{*,c}(nlm)~.\label{GCPAllowedCondition}
\end{equation}
Otherwise, the cloud either cannot form, or would have been depleted via $\tilde{\Gamma}_{nlm}<0$ long before entering the resonance band.

\begin{figure}[h!]
	\centering
	\includegraphics[width=8.5cm]{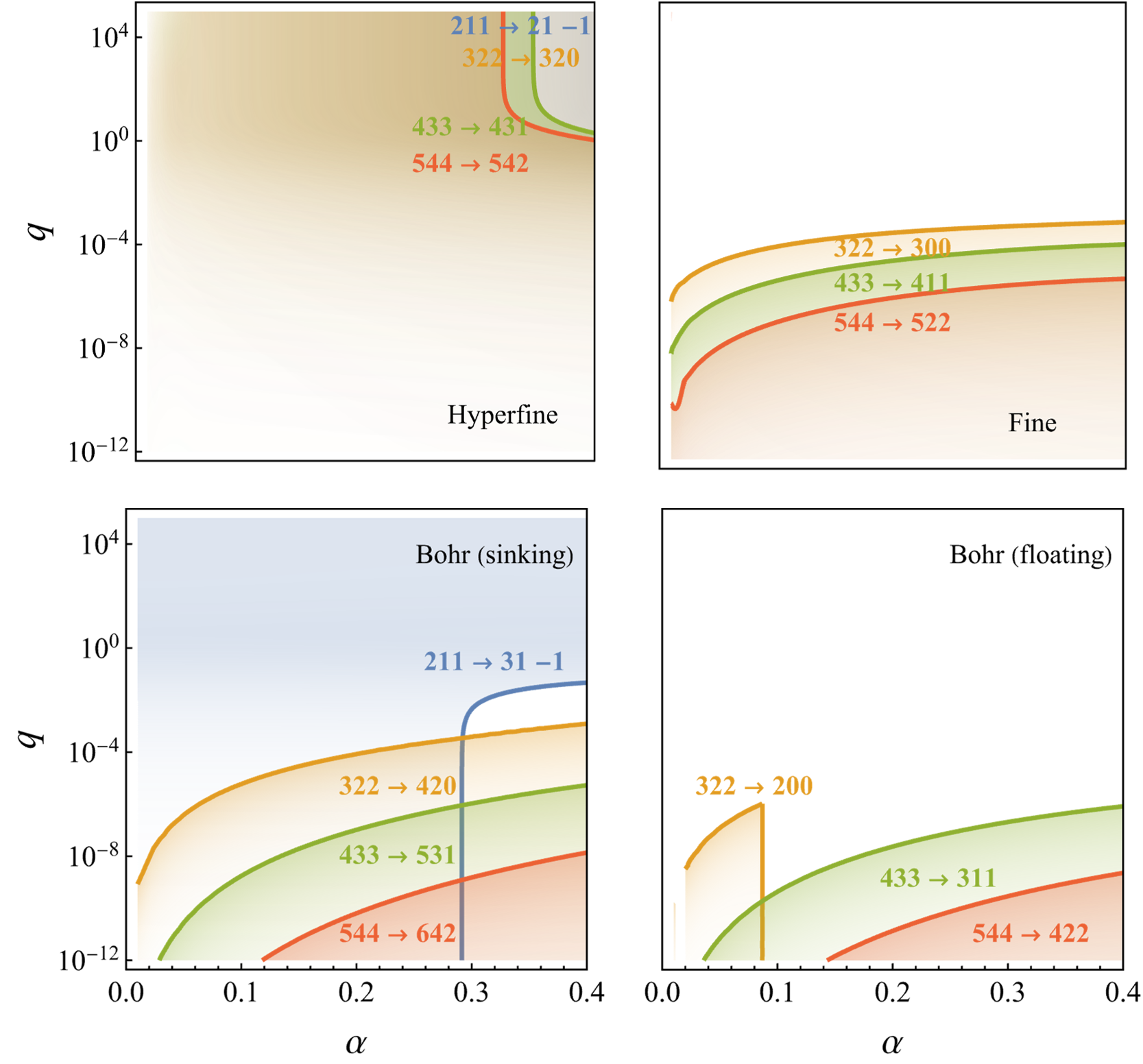}
	\caption{The ``safe'' region for having various hyperfine, fine and Bohr GCP transitions. The solid lines represent the boundary where the resonance distance is equal to the critical distance, and its shaded side satisfies (\ref{GCPAllowedCondition}) and is allowed. Note that the hyperfine transitions $|\psi_{211}\rangle\to|\psi_{21~-1}\rangle$ and $|\psi_{322}\rangle\to|\psi_{300}\rangle$ are allowed throughout the parameter space. For Bohr transitions, we also included the constraint $R_{*,r}(nlm\to n'l'm')>\max \left\{n^2 r_1, n'^2 r_1\right\}$ to keep the validity of tidal perturbation theory. The black hole spin is chosen to be the saturation value of the initial cloud state.}\label{figGCPImpact}
\end{figure}

Scanning through the parameter space, we plot the ``safe'' region for various GCP transitions in FIG.~\ref{figGCPImpact}. It is clear from the plot that superradiance termination poses a tight bound on the parameters of the binary system. Bohr transitions with floating orbits are most severely constrained, because they happen at relatively high orbital frequencies. This means the binary separation at resonance is short, and the cloud is vulnerable to absorption. Hyperfine transitions, on the contrary, are influenced the least. In particular, we note that the $|\psi_{211}\rangle\to|\psi_{21~-1}\rangle$ and $|\psi_{322}\rangle\to|\psi_{300}\rangle$ transitions are safe to occur throughout the parameter space. This is because they happen at low orbital frequencies, where the overlap between the superradiant mode and absorptive modes is still small. Overall, except the hyperfine ones and Bohr transitions starting from the state $|\psi_{211}\rangle$, most transitions are forbidden for mass ratio $q\gtrsim10^{-3}$. They are allowed only below a certain mass ratio, corresponding to a small mass of the binary companion.

\subsection{Observing cloud absorption}\label{OffResObservability}
Before entering the critical distance $R_{*,c}$, the cloud may have already grown up. Then the negative effective superradiance rate within $R_{*,c}$ will gradually deplete the cloud. The loss of angular-momentum-carrying particles backreacts on the orbit and produces an effective torque on the binary companion, leading to floating/sinking orbits in a manner similar to GCP transitions. The difference here is that such orbital backreaction does not occur at a fixed resonance band.

\begin{figure}[h!]
	\centering
	\includegraphics[width=7cm]{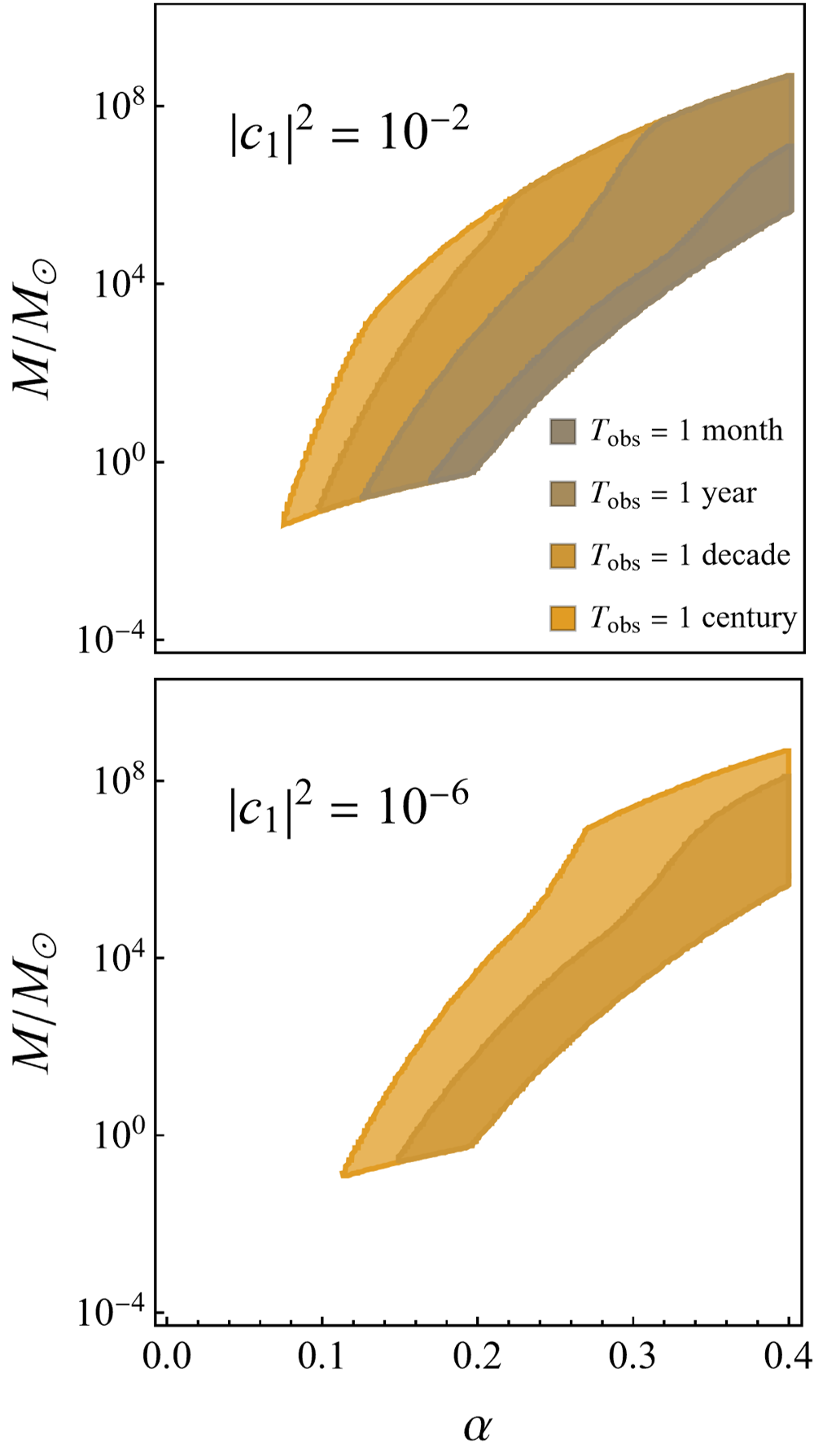}
	\caption{The parameter region reachable for detecting cloud absorption using a pulsar with mass $M_*=1.4 M_{\odot}$ and width $w=0.01$s, for different cloud occupation fraction $|c_1|^2$. Here we have taken the state $|\psi_{322}\rangle$ as an example. The orbital period is chosen to be near the critical distance, $P=0.95 P(R_{*,c}(322))$. Different contours represent different total observational time $T_{\text{obs}}$. The daily observation time is chosen to be $t_{\text{obs}}=5$ hr. In addition to the timing accuracy, requirements on cloud formation and depletion are also included: $T^{(\text{grow})}>10^6$ yr, $T^{(\text{deplete})}<10^8$ yr \cite{baumann2020gravitational,ding2021gravitational}.}\label{figPSR}
\end{figure}

The total angular momentum of the cloud is
\begin{equation}
	S_c(t)=S_{c,0}\left(m_1|c_1(t)|^2+\sum_{i\neq 1} m_i |c_i(t)|^2\right)~,
\end{equation}
where $S_{c,0}$ is the cloud angular momentum at the saturation of $|\psi_1\rangle$, and $0\leqslant|c_1|^2\leqslant1$ is the percentage of cloud occupation. It effectively describes the boson particle number, $i.e.$, the larger $|c_1|^2$ is, there are more particles in the cloud. Since except the superradiant mode $|\psi_1\rangle$ with $m_1>0$, most other modes are usually unoccupied ($c_i\simeq 0$), we can approximate $S_c(t)\simeq S_{c,0}m_1|c_1(t)|^2$. In the case of a planar circular orbit, considering backreaction of the cloud evolution yields an orbital period derivative
\begin{equation}
	\dot{P}=(\dot{P})_{\text{GR}}+(\dot{P})_{\text{C}}~,
\end{equation}
where $(\dot{P})_{\text{GR}}$ represents the usual orbital decay in General Relativity (GR),
\begin{equation}
	(\dot{P})_{\text{GR}}=-\frac{96}{5}(2\pi)^{8/3}\frac{q}{(1+q)^{1/3}}M^{5/3}P^{-5/3}~,
\end{equation}
and $(\dot{P})_{\text{C}}$ is the extra contribution due to cloud backreaction \cite{baumann2020gravitational,ding2021gravitational},
\begin{equation}
	(\dot{P})_{\text{C}}=-3(2\pi)^{1/3}(1+q)^{-2/3}\frac{S_{c,0}m_1}{M^2}\frac{d|c_1(t)|^2}{dt}M^{1/3} P^{2/3}~.
\end{equation}
The cloud evolution would have been unitary but for the absorption, which means $c_1(t)\propto e^{-\tilde\Gamma_1 t}$ and
\begin{equation}
	\frac{d|c_1(t)|^2}{dt}=-2\tilde\Gamma_1(R_{*}) |c_1(t)|^2~.
\end{equation}
For instance, using Kepler's law to rewrite $R_{*}$ in terms of $P$, we can obtain the fractional correction to the period derivative for $|\psi_{322}\rangle$,
\begin{align}
\nonumber&\frac{(\dot{P})_{\text{C}}}{(\dot{P})_{\text{GR}}}\\
\simeq&-15|c_{322}|^2\left(\frac{\alpha}{0.1}\right)^{-9}\frac{q}{(1+q)^{7/3}}\left(\frac{M}{10 M_{\odot}}\right)^{5/3}\left(\frac{P}{1\text{ hr}}\right)^{-5/3}.
\end{align}
The minus sign shows that this is a floating orbit. Thus the correction to the orbital period derivative can be significant for certain parameter choices\footnote{Since the cloud has been decaying for some time after entering the critical distance, we generally expect $|c_1|^2$ to be a small number. In principle, assuming natural evolution, one can solve the whole history and determine $|c_1|^2$. Here, for simplicity, we will treat it as a free parameter.}. Such corrections should be detectable via multi-messengers such as gravitational waves and pulsar timing \cite{Hulse:1974eb,Weisberg:2004hi}. For instance, consider the gravitational atom in a pulsar-black hole binary, the periastron time shift is calculated by
\begin{equation}
	\Delta_P\equiv t-P(0)\int_0^t\frac{dt'}{P(t')}\approx \frac{1}{2}\frac{\dot{P}}{P}t^2~,
\end{equation}
where we have Taylor expanded the orbital period to the linear order. In order to observe the extra periastron time shift caused by the backreaction of cloud absorption, we must require the deviation from the GR result to be larger than the timing error of pulse counting,
\begin{equation}
	|\Delta_P-(\Delta_P)_{\text{GR}}|>\sigma_{\Delta_P}~.
\end{equation}
For a pulsar with rotation period $\tau$ and pulse width $w\lesssim\tau$, the timing error $\sigma_{\Delta_P}$ can be roughly estimated as \cite{ding2021gravitational}
\begin{equation}
	\sigma_{\Delta_P}\sim\frac{1}{\sqrt{t/1 \text{day}}}\frac{w}{t_{\text{obs}}/P}~, \label{PeriUncertainty}
\end{equation}
where $t_{obs}$ is the duration of a continuous observation window every day. In FIG.~\ref{figPSR}, we plot the timing accuracy for a pulsar of width $w=0.01$s and total observation time $T_{\text{obs}}$. We see that with enough observational time ($e.g.$, $T_{\text{obs}}\sim$ 1 decade), much parameter region can be uncovered, even when the cloud is extremely dilute ($e.g.$, $|c_1|^2\sim 10^{-6}$).

\subsection{Toward relaxing the bound on boson mass}
As mentioned before, boson clouds in isolated black holes only deplete by emitting gravitational waves whose frequency is peaked around the boson rest mass $\mu$. Thus the null detection of such near-monochromatic gravitational waves poses constraints on the mass of the ultralight boson, under certain assumptions on the initial black hole spin, cloud ages and the distance from earth. For instance, the LIGO search for continuous gravitational waves from the Milky Way center has reported constraints for the boson mass rage $10^{-13}$eV$<\mu<10^{-12}$eV \cite{LIGOScientific:2022lsr} (see also \cite{Palomba:2019vxe,Ng:2020jqd,LIGOScientific:2021jlr,Yuan:2022bem}). However, it is likely that spinning black holes are \textit{not} isolated in the galactic center, where the density of objects is high. Therefore, in such a high-density and highly dynamical region, black hole superradiance may be influenced by the presence of a binary companion or other objects, possibly even suppressed to the extent of terminated cloud growth. Henceforth, the gravitational wave null detection bound could be relaxed. We have also seen in Sect.~\ref{FullCDSection} that states with higher $l$ are more severely affected by the tidal perturbation of nearby objects. Thus we expect the bound should be state-dependent, with cloud states of higher $l$ being constrained less.

Another boson mass bound comes from measuring the mass-spin distribution of black holes (the so-called Regge plane). Assuming efficient superradiance growth of the cloud, the black hole angular momentum will quickly be extracted, and its spin parameter is cut off at a value less than 1. This effect manifests itself as a gap on the black hole Regge plane, which can be tested statistically with LIGO and LISA \cite{Arvanitaki:2016qwi,Brito:2017zvb,Ng:2020ruv}. Considering superradiance termination, such constraints are also subjected to relaxation if the black hole is in a binary. As mentioned in Sect.~\ref{FullCDSection}, if the superradiance of a certain state is shut off by a close companion, the black hole spin cannot further decrease by producing that state. If, in addition, this superradiant state is the one with the lowest $l$ that the black hole spin is capable of producing, all other states will be shut off. For example, if $R_*<R_{*,c}(322)$, and $\tilde{a}<\frac{4\alpha}{1+4\alpha^2}$, both $|\psi_{322}\rangle$ and $|\psi_{211}\rangle$ cannot grow, and neither can the higher-$l$ modes (since they have even larger $R_{*,c}$). Then the black hole spin will not be cut off at the saturation value $\frac{\alpha}{1+\alpha^2}$ of the $|\psi_{322}\rangle$ state. Such considerations must be taken into account in the statistical analysis of the black hole Regge plane.

Other bounds come from, for example, testing the birefringence of photons \cite{Chen:2019fsq} and the stellar kinematics \cite{Yuan:2022nmu} near supermassive black holes, with the help of powerful telescopes such as EHT \cite{EventHorizonTelescope:2019dse,EventHorizonTelescope:2022xnr} and the Keck Observatory \cite{Do:2019txf}. These bounds also assume the presence of cloud, and are subjected to a similar relaxation as mentioned here, since supermassive black holes are known to be surrounded by numerous stars. A detailed analysis of the impact of superradiance termination on various boson mass bounds is beyond the scoop of this paper, but it no doubt deserves further investigations in future works.

\section{Conclusion}\label{ConclusionSection}
Superradiance instability of ultralight bosons near a rotating black hole leads to lots of interesting phenomena, many of which are based on the existence of a hydrogen-atom-like cloud. In this work, however, we raise the question on the robustness of black hole superradiance in the presence of a binary companion. We have found that through the tidal perturbation of the companion, superradiant states can be coupled to dangerous absorptive states and receive negative corrections to its growth rate, thereby suppressing superradiance. We have found that for a given cloud state, there exists a critical binary distance below which its superradiance is terminated. This fact leads to several important consequences. On one hand, it poses tight constraints on possible GCP transitions. For instance, except hyperfine transitions and Bohr transitions starting from $|\psi_{211}\rangle$, almost all other transitions must have a mass ratio $q\ll 1$. On the other hand, after entering the critical distance, an existing cloud is absorbed back into the black hole. This process can be observed via multiple messengers such as pulsar timing. In addition, the termination of superradiance implies the absence of certain cloud states. This effect is expected to relax the constraints on the ultralight boson mass using various methods that rely on the existence of such clouds.

We have followed a simplistic route in the current work. And there are certainly many improvements to make and prospects to explore in the future. To name a few, first, the analysis of superradiance termination itself can be generalized to more realistic scenarios with generic orbits. Second, we need to understand in detail how superradiance termination can affect the current boson mass bounds. Third, aside from bounded binary orbits considered here, external objects with unbounded orbits may also influence the cloud. It is then interesting to study the stability of the gravitational atom for more complex dynamics such as three-body evolutions and scattering with stellar objects.

~\\

\section*{Acknowledgments}
We thank Chon Man Sou for helpful discussions. XT and YW were supported in part by the National Key R\&D Program of China (2021YFC2203100),
the NSFC Excellent Young Scientist Scheme (Hong Kong and Macau) Grant No. 12022516, and by the RGC of Hong Kong SAR, China (Project No.~CRF C6017-20GF). HYZ was supported in part by a grant from the RGC of the Hong Kong SAR, China (Project No.~16303220).
\\

\bibliographystyle{utphys}
\bibliography{ref}

\providecommand{\href}[2]{#2}\begingroup\raggedright\begin{thebibliography}{10}

\bibitem{zel1971generation}
Y.~B. Zel'Dovich, ``Generation of waves by a rotating body,'' {\em Soviet
  Journal of Experimental and Theoretical Physics Letters} {\bfseries 14}
  (1971) 180.

\bibitem{zel1972amplification}
I.~ZEL'DOVICH, ``Amplification of cylindrical electromagnetic waves reflected
  from a rotating body,'' {\em Soviet Physics-JETP} {\bfseries 35} (1972)
  1085--1087.

\bibitem{brito2020superradiance}
R.~Brito, V.~Cardoso, and P.~Pani, ``{Superradiance}: {New Frontiers in Black
  Hole Physics},'' \href{http://dx.doi.org/10.1007/978-3-319-19000-6}{{\em
  Lect. Notes Phys.} {\bfseries 906} (2015) pp.1--237},
  \href{http://arxiv.org/abs/1501.06570}{{\ttfamily arXiv:1501.06570 [gr-qc]}}.

\bibitem{Press:1972zz}
W.~H. Press and S.~A. Teukolsky, ``{Floating Orbits, Superradiant Scattering
  and the Black-hole Bomb},'' \href{http://dx.doi.org/10.1038/238211a0}{{\em
  Nature} {\bfseries 238} (1972) 211--212}.

\bibitem{Damour:1976kh}
T.~Damour, N.~Deruelle, and R.~Ruffini, ``{On Quantum Resonances in Stationary
  Geometries},'' \href{http://dx.doi.org/10.1007/BF02725534}{{\em Lett. Nuovo
  Cim.} {\bfseries 15} (1976) 257--262}.

\bibitem{arvanitaki2010string}
A.~Arvanitaki, S.~Dimopoulos, S.~Dubovsky, N.~Kaloper, and J.~March-Russell,
  ``{String Axiverse},''
  \href{http://dx.doi.org/10.1103/PhysRevD.81.123530}{{\em Phys. Rev. D}
  {\bfseries 81} (2010) 123530},
  \href{http://arxiv.org/abs/0905.4720}{{\ttfamily arXiv:0905.4720 [hep-th]}}.

\bibitem{arvanitaki2011exploring}
A.~Arvanitaki and S.~Dubovsky, ``{Exploring the String Axiverse with Precision
  Black Hole Physics},''
  \href{http://dx.doi.org/10.1103/PhysRevD.83.044026}{{\em Phys. Rev. D}
  {\bfseries 83} (2011) 044026},
  \href{http://arxiv.org/abs/1004.3558}{{\ttfamily arXiv:1004.3558 [hep-th]}}.

\bibitem{Arvanitaki:2016qwi}
A.~Arvanitaki, M.~Baryakhtar, S.~Dimopoulos, S.~Dubovsky, and R.~Lasenby,
  ``{Black Hole Mergers and the QCD Axion at Advanced LIGO},''
  \href{http://dx.doi.org/10.1103/PhysRevD.95.043001}{{\em Phys. Rev. D}
  {\bfseries 95} no.~4, (2017) 043001},
  \href{http://arxiv.org/abs/1604.03958}{{\ttfamily arXiv:1604.03958
  [hep-ph]}}.

\bibitem{Brito:2017zvb}
R.~Brito, S.~Ghosh, E.~Barausse, E.~Berti, V.~Cardoso, I.~Dvorkin, A.~Klein,
  and P.~Pani, ``{Gravitational wave searches for ultralight bosons with LIGO
  and LISA},'' \href{http://dx.doi.org/10.1103/PhysRevD.96.064050}{{\em Phys.
  Rev. D} {\bfseries 96} no.~6, (2017) 064050},
  \href{http://arxiv.org/abs/1706.06311}{{\ttfamily arXiv:1706.06311 [gr-qc]}}.

\bibitem{Ng:2020ruv}
K.~K.~Y. Ng, S.~Vitale, O.~A. Hannuksela, and T.~G.~F. Li, ``{Constraints on
  Ultralight Scalar Bosons within Black Hole Spin Measurements from the
  LIGO-Virgo GWTC-2},''
  \href{http://dx.doi.org/10.1103/PhysRevLett.126.151102}{{\em Phys. Rev.
  Lett.} {\bfseries 126} no.~15, (2021) 151102},
  \href{http://arxiv.org/abs/2011.06010}{{\ttfamily arXiv:2011.06010 [gr-qc]}}.

\bibitem{yoshino2014gravitational}
H.~Yoshino and H.~Kodama, ``{Gravitational radiation from an axion cloud around
  a black hole: Superradiant phase},''
  \href{http://dx.doi.org/10.1093/ptep/ptu029}{{\em PTEP} {\bfseries 2014}
  (2014) 043E02}, \href{http://arxiv.org/abs/1312.2326}{{\ttfamily
  arXiv:1312.2326 [gr-qc]}}.

\bibitem{Palomba:2019vxe}
C.~Palomba {\em et~al.}, ``{Direct constraints on ultra-light boson mass from
  searches for continuous gravitational waves},''
  \href{http://dx.doi.org/10.1103/PhysRevLett.123.171101}{{\em Phys. Rev.
  Lett.} {\bfseries 123} (2019) 171101},
  \href{http://arxiv.org/abs/1909.08854}{{\ttfamily arXiv:1909.08854
  [astro-ph.HE]}}.

\bibitem{LIGOScientific:2021jlr}
{\bfseries LIGO Scientific, VIRGO, KAGRA} Collaboration, R.~Abbott {\em
  et~al.}, ``{All-sky search for gravitational wave emission from scalar boson
  clouds around spinning black holes in LIGO O3 data},''
  \href{http://arxiv.org/abs/2111.15507}{{\ttfamily arXiv:2111.15507
  [astro-ph.HE]}}.

\bibitem{LIGOScientific:2022lsr}
{\bfseries LIGO Scientific, VIRGO, KAGRA} Collaboration, R.~Abbott {\em
  et~al.}, ``{Search for continuous gravitational wave emission from the Milky
  Way center in O3 LIGO--Virgo data},''
  \href{http://arxiv.org/abs/2204.04523}{{\ttfamily arXiv:2204.04523
  [astro-ph.HE]}}.

\bibitem{Yuan:2022bem}
C.~Yuan, Y.~Jiang, and Q.-G. Huang, ``{Constraints on the ultralight scalar
  boson from Advanced LIGO and Advanced Virgo's first three observing runs
  using the stochastic gravitational-wave background},''
  \href{http://arxiv.org/abs/2204.03482}{{\ttfamily arXiv:2204.03482
  [astro-ph.CO]}}.

\bibitem{baumann2019probing}
D.~Baumann, H.~S. Chia, and R.~A. Porto, ``{Probing Ultralight Bosons with
  Binary Black Holes},''
  \href{http://dx.doi.org/10.1103/PhysRevD.99.044001}{{\em Phys. Rev. D}
  {\bfseries 99} no.~4, (2019) 044001},
  \href{http://arxiv.org/abs/1804.03208}{{\ttfamily arXiv:1804.03208 [gr-qc]}}.

\bibitem{baumann2020gravitational}
D.~Baumann, H.~S. Chia, R.~A. Porto, and J.~Stout, ``{Gravitational Collider
  Physics},'' \href{http://dx.doi.org/10.1103/PhysRevD.101.083019}{{\em Phys.
  Rev. D} {\bfseries 101} no.~8, (2020) 083019},
  \href{http://arxiv.org/abs/1912.04932}{{\ttfamily arXiv:1912.04932 [gr-qc]}}.

\bibitem{Baumann:2021fkf}
D.~Baumann, G.~Bertone, J.~Stout, and G.~M. Tomaselli, ``{Ionization of
  Gravitational Atoms},'' \href{http://arxiv.org/abs/2112.14777}{{\ttfamily
  arXiv:2112.14777 [gr-qc]}}.

\bibitem{ding2021gravitational}
Q.~Ding, X.~Tong, and Y.~Wang, ``{Gravitational Collider Physics via
  Pulsar-Black Hole Binaries},''
  \href{http://dx.doi.org/10.3847/1538-4357/abd803}{{\em Astrophys. J.}
  {\bfseries 908} no.~1, (2021) 78},
  \href{http://arxiv.org/abs/2009.11106}{{\ttfamily arXiv:2009.11106
  [astro-ph.HE]}}.

\bibitem{tong2022gravitational}
X.~Tong, Y.~Wang, and H.-Y. Zhu, ``{Gravitational Collider Physics via
  Pulsar\textendash{}Black Hole Binaries II: Fine and Hyperfine Structures Are
  Favored},'' \href{http://dx.doi.org/10.3847/1538-4357/ac36db}{{\em Astrophys.
  J.} {\bfseries 924} no.~2, (2022) 99},
  \href{http://arxiv.org/abs/2106.13484}{{\ttfamily arXiv:2106.13484
  [astro-ph.HE]}}.

\bibitem{su2021probing}
B.~Su, Z.-Z. Xianyu, and X.~Zhang, ``{Probing Ultralight Bosons with Compact
  Eccentric Binaries},'' \href{http://dx.doi.org/10.3847/1538-4357/ac2d91}{{\em
  Astrophys. J.} {\bfseries 923} no.~1, (2021) 114},
  \href{http://arxiv.org/abs/2107.13527}{{\ttfamily arXiv:2107.13527 [gr-qc]}}.

\bibitem{Ikeda:2020xvt}
T.~Ikeda, L.~Bernard, V.~Cardoso, and M.~Zilh\~ao, ``{Black hole binaries and
  light fields: Gravitational molecules},''
  \href{http://dx.doi.org/10.1103/PhysRevD.103.024020}{{\em Phys. Rev. D}
  {\bfseries 103} no.~2, (2021) 024020},
  \href{http://arxiv.org/abs/2010.00008}{{\ttfamily arXiv:2010.00008 [gr-qc]}}.

\bibitem{Ficarra:2021qeh}
G.~Ficarra, ``{Scalar field dynamics around black holes: superradiant
  instabilities and binary evolution},'' in {\em {55th Rencontres de Moriond on
  Gravitation}}.
\newblock 5, 2021.
\newblock \href{http://arxiv.org/abs/2105.05918}{{\ttfamily arXiv:2105.05918
  [gr-qc]}}.

\bibitem{liu2021bh}
T.~Liu and K.-F. Lyu, ``{The BH-PSR Gravitational Molecule},''
  \href{http://arxiv.org/abs/2107.09971}{{\ttfamily arXiv:2107.09971
  [astro-ph.HE]}}.

\bibitem{Berti:2019wnn}
E.~Berti, R.~Brito, C.~F.~B. Macedo, G.~Raposo, and J.~L. Rosa, ``{Ultralight
  boson cloud depletion in binary systems},''
  \href{http://dx.doi.org/10.1103/PhysRevD.99.104039}{{\em Phys. Rev. D}
  {\bfseries 99} no.~10, (2019) 104039},
  \href{http://arxiv.org/abs/1904.03131}{{\ttfamily arXiv:1904.03131 [gr-qc]}}.

\bibitem{Takahashi:2021eso}
T.~Takahashi and T.~Tanaka, ``{Axion clouds may survive the perturbative tidal
  interaction over the early inspiral phase of black hole binaries},''
  \href{http://dx.doi.org/10.1088/1475-7516/2021/10/031}{{\em JCAP} {\bfseries
  10} (2021) 031}, \href{http://arxiv.org/abs/2106.08836}{{\ttfamily
  arXiv:2106.08836 [gr-qc]}}.

\bibitem{Wong:2020qom}
L.~K. Wong, ``{Evolution of diffuse scalar clouds around binary black holes},''
  \href{http://dx.doi.org/10.1103/PhysRevD.101.124049}{{\em Phys. Rev. D}
  {\bfseries 101} no.~12, (2020) 124049},
  \href{http://arxiv.org/abs/2004.03570}{{\ttfamily arXiv:2004.03570
  [hep-th]}}.

\bibitem{Takahashi:2021yhy}
T.~Takahashi, H.~Omiya, and T.~Tanaka, ``{Axion cloud evaporation during
  inspiral of black hole binaries -- the effects of backreaction and
  radiation},'' \href{http://arxiv.org/abs/2112.05774}{{\ttfamily
  arXiv:2112.05774 [gr-qc]}}.

\bibitem{baumann2019spectra}
D.~Baumann, H.~S. Chia, J.~Stout, and L.~ter Haar, ``{The Spectra of
  Gravitational Atoms},''
  \href{http://dx.doi.org/10.1088/1475-7516/2019/12/006}{{\em JCAP} {\bfseries
  12} (2019) 006}, \href{http://arxiv.org/abs/1908.10370}{{\ttfamily
  arXiv:1908.10370 [gr-qc]}}.

\bibitem{bao2022improved}
S.~Bao, Q.-X. Xu, and H.~Zhang, ``{Improved Analytic Solution of Black Hole
  Superradiance},'' \href{http://arxiv.org/abs/2201.10941}{{\ttfamily
  arXiv:2201.10941 [gr-qc]}}.

\bibitem{detweiler1980klein}
S.~L. Detweiler, ``{Klein-Gordon equation and rotating black holes},''
  \href{http://dx.doi.org/10.1103/PhysRevD.22.2323}{{\em Phys. Rev. D}
  {\bfseries 22} (1980) 2323--2326}.

\bibitem{takagi1991quantum}
S.~Takagi, ``{Quantum Dynamics and Non-Inertial Frames of Reference. I:
  Generality},'' {\em Progress of theoretical physics} {\bfseries 85} no.~3,
  (1991) 463--479.

\bibitem{Hulse:1974eb}
R.~A. Hulse and J.~H. Taylor, ``{Discovery of a pulsar in a binary system},''
  \href{http://dx.doi.org/10.1086/181708}{{\em Astrophys. J. Lett.} {\bfseries
  195} (1975) L51--L53}.

\bibitem{Weisberg:2004hi}
J.~M. Weisberg and J.~H. Taylor, ``{Relativistic binary pulsar B1913+16: Thirty
  years of observations and analysis},'' {\em ASP Conf. Ser.} {\bfseries 328}
  (2005) 25, \href{http://arxiv.org/abs/astro-ph/0407149}{{\ttfamily
  arXiv:astro-ph/0407149}}.

\bibitem{Ng:2020jqd}
K.~K.~Y. Ng, M.~Isi, C.-J. Haster, and S.~Vitale, ``{Multiband
  gravitational-wave searches for ultralight bosons},''
  \href{http://dx.doi.org/10.1103/PhysRevD.102.083020}{{\em Phys. Rev. D}
  {\bfseries 102} no.~8, (2020) 083020},
  \href{http://arxiv.org/abs/2007.12793}{{\ttfamily arXiv:2007.12793 [gr-qc]}}.

\bibitem{Chen:2019fsq}
Y.~Chen, J.~Shu, X.~Xue, Q.~Yuan, and Y.~Zhao, ``{Probing Axions with Event
  Horizon Telescope Polarimetric Measurements},''
  \href{http://dx.doi.org/10.1103/PhysRevLett.124.061102}{{\em Phys. Rev.
  Lett.} {\bfseries 124} no.~6, (2020) 061102},
  \href{http://arxiv.org/abs/1905.02213}{{\ttfamily arXiv:1905.02213
  [hep-ph]}}.

\bibitem{Yuan:2022nmu}
G.-W. Yuan, Z.-Q. Shen, Y.-L.~S. Tsai, Q.~Yuan, and Y.-Z. Fan, ``{Constraining
  ultralight bosonic dark matter with Keck observations of S2's orbit and
  kinematics},'' \href{http://arxiv.org/abs/2205.04970}{{\ttfamily
  arXiv:2205.04970 [astro-ph.HE]}}.

\bibitem{EventHorizonTelescope:2019dse}
{\bfseries Event Horizon Telescope} Collaboration, K.~Akiyama {\em et~al.},
  ``{First M87 Event Horizon Telescope Results. I. The Shadow of the
  Supermassive Black Hole},''
  \href{http://dx.doi.org/10.3847/2041-8213/ab0ec7}{{\em Astrophys. J. Lett.}
  {\bfseries 875} (2019) L1}, \href{http://arxiv.org/abs/1906.11238}{{\ttfamily
  arXiv:1906.11238 [astro-ph.GA]}}.

\bibitem{EventHorizonTelescope:2022xnr}
{\bfseries Event Horizon Telescope} Collaboration, K.~Akiyama {\em et~al.},
  ``{First Sagittarius A* Event Horizon Telescope Results. I. The Shadow of the
  Supermassive Black Hole in the Center of the Milky Way},''
  \href{http://dx.doi.org/10.3847/2041-8213/ac6674}{{\em Astrophys. J. Lett.}
  {\bfseries 930} no.~2, (2022) L12}.

\bibitem{Do:2019txf}
T.~Do {\em et~al.}, ``{Relativistic redshift of the star S0-2 orbiting the
  Galactic center supermassive black hole},''
  \href{http://dx.doi.org/10.1126/science.aav8137}{{\em Science} {\bfseries
  365} no.~6454, (2019) 664--668},
  \href{http://arxiv.org/abs/1907.10731}{{\ttfamily arXiv:1907.10731
  [astro-ph.GA]}}.

\end{thebibliography}\endgroup

\end{document}